\documentclass[aps,prX,twocolumn,superscriptaddress,longbibliography]{revtex4-1}

\usepackage{fancyhdr}     
\usepackage{amsfonts}
\usepackage{amsmath}
\usepackage[latin1]{inputenc}
\usepackage{graphicx,graphics,amssymb,times}
\usepackage{color}
\usepackage{natbib}
\usepackage{comment}
\usepackage{mathtools}
\usepackage[normalem]{ulem}
\usepackage{array}
\usepackage{siunitx}
\usepackage{array}
\usepackage{makecell}
\usepackage{booktabs}
\usepackage{multirow}
\usepackage{ragged2e}
\usepackage[table]{xcolor}
\usepackage[thinlines]{easytable}
\newcolumntype{P}[1]{>{\centering\arraybackslash}p{#1}}

\begin{document}


\title{Optical computing of quantum revivals}

\begin{abstract}
Interference is the mechanism through which waves can be structured into the most fascinating patterns. While for sensing, imaging, trapping, or in fundamental investigations, structured waves play nowadays an important role and are becoming subject of many interesting studies. Using a coherent optical field as a probe, we show how to structure light into distributions presenting collapse and revival structures in its wavefront. These distributions are obtained from the Fourier spectrum of an arrangement of aperiodic diffracting structures. Interestingly, the resulting interference may present quasiperiodic structures of diffraction peaks on a number of distance scales, even though the diffracting structure is not periodic. We establish an analogy with revival phenomena in the evolution of quantum mechanical systems and illustrate this computation numerically and experimentally, obtaining excellent agreement with the proposed theory. 
\end{abstract}

\author{M. R. Maia} \email[electronic address: ]{mayanne.maia@iff.edu.br}
\affiliation{Instituto de F\'isica, Universidade Federal Fluminense, Niter\'{o}i, RJ
24210-346, Brazil}
\affiliation{Instituto Federal Fluminense, Bom Jesus do Itabapoana, RJ 28360-000, Brazil}
\author{D. Jonathan}
\author{T. R. de Oliveira}
\author{A. Z. Khoury}
\author{D. S. Tasca} \email[electronic address: ]{danieltasca@id.uff.br}
\affiliation{Instituto de F\'isica, Universidade Federal Fluminense, Niter\'{o}i, RJ
24210-346, Brazil}

\maketitle

\date{\today}

\section{Introduction} 
Interference patterns arising from diffraction of light in multiple apertures date back from the beginning of the nineteenth century, when Thomas Young demonstrated the wave-like behaviour of light through his famous double slit experiment. The corresponding experiment using single electrons has been called \textit{the most beautiful experiment in physics} \cite{Crease02}. Matter wave interference has also been demonstrated with fullerene \cite{Arndt99} and even heavier molecules \cite{Juffmann12}, but this time using periodic gratings. In optics, diffraction from periodic gratings plays an important role in many applications, such as spectrometry and linewidth control of laser beams \cite{Bonod:2016tk}. The interference pattern from periodic structures is also an important mechanism for material characterisation such as in crystallography, both using matter (electron, neutron) as well as electromagnetic (X-ray) waves.

Much less attention has been given to \textit{aperiodic} arrangements of diffracting structures. In X-ray diffraction, for instance, completely disordered or amorphous media do not give rise to sharp diffraction peaks, while aperiodic lattices with long-range order (quasi-crystals) manifest diffraction signatures similar to periodic lattices \cite{Grimm15,Strzalka16}. Interference from aperiodic structures has been studied in the context of photonic and plasmonic quasi-crystals \cite{Freedman06,Zotov10,Dal-Negro12} as well as in matter waves interacting with quasiperiodic optical lattices  \cite{Sanchez-Palencia:2005,Viebahn:2019}.
Besides, studies on the focusing and deflection properties of aperiodic gratings have been carried out \cite{Mata-Mendez2008,Das2013,Ma2014,Youssef2016,Torcal-Milla2008A,Torcal-Milla2008B}, with attention to structures ruled by Fibonacci sequences  \cite{Verma2014,Verma2014b,Gullo2017}.

In this work, we investigate diffracting structures giving rise to interference patterns displaying interesting phenomena such as \textit{collapses} and \textit{revivals} of the far-field diffraction distribution. Our diffraction grating is devised as an array of slits positioned aperiodically along a single dimension, transverse to the beam propagation axis. On one hand, the grating Fourier spectrum presents sharp peaks of interference maxima, similar to that of a periodic one. On the other hand, the peak distribution is structured into patterns presenting quasi-periodic oscillations on distance scales differing by orders of magnitude from one another. We show how to structure patterns of diffraction peaks closely grouped into sectors, each of which spaced from one another by dark regions that can be made arbitrarily long by adjustment of the grating parameters and/or illumination pattern. 
These dark regions correspond to `collapsed' detection probability, caused by nearly complete destructive interference over long distances in the detection plane. Interestingly, the length of these dark regions is interpreted in perfect analogy with revival times of expectation values corresponding to quantum mechanical observables. Whilst the quantised nature of the quantum system is reproduced by the thin slits of the array, their aperiodic disposition allows us in principle to model the discrete portion of the spectrum of any Hamiltonian. This analogy allows us to compute interesting patterns of quantum revivals using a simple, yet rich optics experiment. It should perhaps be emphasised that this does \emph{not} imply that our setup can efficiently compute arbitrary properties of a quantum system - a task that requires a universal quantum computer \cite{Nielsen2000}. Indeed, it is well-known \cite{Kok2007} that linear optical setups such as ours cannot perform universal quantum computation, except if supplemented by nonlinear elements such as single-photon sources, nonlinear media, or photon counters.

\section{Methods}

In our experiment, we observe the far-field distribution of the diffracted light field by means of a Fourier transforming lens.  We begin therefore by reviewing in this section some relevant aspects of Fraunhofer diffraction. 

\subsection{Optical Fourier Transform}
For simplicity, we assume an input field distribution that is factorisable over the orthogonal transverse coordinates $x$ and $y$, where $x$ is the direction along which the slits are distributed at the front focal plane of the lens. It therefore suffices
to describe the structure $\psi(x)$ of the field distribution along this direction. After propagation through the optical system, the transformed field reads
\begin{equation}\label{OpticalFourier}
 \tilde{\psi}({x^\prime}) =\sqrt{ \frac{\xi}{2\pi }} \int dx \, \psi(x) e^{-i x (\xi {x^\prime})},
\end{equation}
where $x'$ denotes the transverse coordinate at the back focal plane and $\xi=2\pi/f\lambda$ is a constant with dimension of length$^{-2}$, with $f$ being the focal length of the lens and $\lambda$ the wavelength of the light field. Thus, the far-field distribution $\tilde{\psi}({x^\prime})$ yields the Fourier spectrum of the input light field with its --spatial-- angular frequencies `$\xi {x^\prime}$' proportional to the transverse coordinate ${x^\prime}$ at the back focal plane of the lens. Throughout this paper, the use of the \textit{tilde} will designate the optically Fourier transformed function according to Eq. \eqref{OpticalFourier}.

\subsection{Fraunhofer diffraction pattern from slit arrays}
We shall make use of an array of slits modelled by rectangular functions:
\begin{equation} 
\label{Slit} 
\Pi(x )=\left\{ \begin{array}{ccc}   1, & \,   |x|   \leq   s/2 \\   0, &  {\rm otherwise}  \end{array} \right. .
\end{equation} 
The function \eqref{Slit} thus model a rectangular slit of width `$s$' centred at $x=0$. The geometry of a general array of slits is then uniquely determined by a set of positions $x_n$ locating the slits,
\begin{equation} 
\label{SlitPos} 
x_n= x_0 +q_n,
\end{equation} 
where $q_n$ is a function of the label $n \in \mathbb{Z}$ numbering each slit of the array: $\Pi_n(x) \coloneqq \Pi(x-x_n)$. Without loss of generality, we assume that $q_0=0$, such that the slit numbered as $n=0$ is located at $x_0$ (a displacement parameter of the whole array). Also, it suffices to consider monotonically increasing functions $q_n$, such that $x_{n+1}>x_n$ $\forall$ $n$. Lastly, we notice that mirror-symmetric (around $x_0$) arrays shall be produced by odd $q_n$ functions: $q_n=-q_{-n}$.

In describing the far-field diffraction of spatially coherent light going through the array, it is useful to employ the Fourier  transform of the slit function given in Eq. \eqref{Slit}. We shall denote $\tilde{\Pi}_n({x^\prime})$ as the \textit{optically} Fourier transformed --such as in Eq. \eqref{OpticalFourier}-- slit function
\begin{equation} 
\label{SlitFourierShift} 
\tilde{\Pi}_n({x^\prime})= \tilde{\Pi}_0({x^\prime}) e^{-i \kappa_n  {x^\prime}},
\end{equation} 
where
\begin{equation} 
\label{kappan} 
\kappa_n \coloneqq \xi q_n =\frac{2\pi}{\lambda f} q_n,
\end{equation} 
and
\begin{equation} 
\label{SlitFourier} 
\tilde{\Pi}_0({x^\prime})= s \sqrt{ \frac{\xi}{2\pi }}   \mathrm{sinc} \left( \frac{\xi{x^\prime} s}{2} \right) e^{-i \xi x_0  {x^\prime}}.
\end{equation} 

Upon controlled illumination of the constructed array, we produce a structured light field behind the slits (at the front focal plane of the lens) given by
\begin{equation} 
\label{Psigeneral} 
\psi(x)= \sum_{n\in \mathbb{Z}} \psi_n (x)=\sum_{n\in \mathbb{Z}} A_n \Pi_n(x),
\end{equation} 
where $\psi_n (x) \coloneqq A_n \Pi_n(x)$ represents the light field behind slit $n$ and we assume its complex amplitude, $A_n$, being uniform over the slit. Note that the field amplitudes $A_n$ may be adjusted by previously modulating the input wavefront and/or by controlling the transmittance function of the slits. 
Making use of Eqs. \eqref{SlitFourierShift} and \eqref{SlitFourier} we can easily write the far-field diffraction pattern (at the back focal plane of the lens) associated with the input field given in Eq. \eqref{Psigeneral}:
\begin{eqnarray} 
\label{DiffGeneral} 
\tilde{\psi}({x^\prime})= \sum_{n\in \mathbb{Z}} \tilde{\psi}_n ({x^\prime})= \tilde{\Pi}_0({x^\prime}) \sum_{n\in \mathbb{Z}} A_n e^{-i \kappa_n  {x^\prime}},
\end{eqnarray} 
with $\kappa_n$ defined in \eqref{kappan}.

Expression \eqref{DiffGeneral}, which describes the Fraunhofer diffraction of a general slit array, displays a typical feature associated with far-field interference patterns: the appearance of \textit{diffraction orders} imposed by the Fourier transform of the slit function. 
However, as we shall see in Section \ref{sec:Results}, when dealing with \textit{aperiodic} arrays the diffraction orders are \textit{not}, in general, the most restrictive modulation of the interference maxima. Other more restrictive envelope modulations may be identified depending on the structure of the array, causing interference maxima to collapse for long regions in the detection plane. One of the main goals of this work is to understand how to control this further modulation structure.

\subsubsection{Periodic arrays} \label{subsec:periodic}
For periodic arrays, the function determining the slit positions is simply linear in $n$ and may be written as $q_n=nT$, where $T$ is the spatial periodicity of the grating. The far-field amplitude then reads
\begin{eqnarray} 
\label{DiffPer} 
\tilde{\psi}_{\mathrm{per}}({x^\prime})= \tilde{\Pi}_0({x^\prime}) \sum_{n\in \mathbb{Z}} A_n e^{-i n T \xi  {x^\prime}}.
\end{eqnarray} 
As is well known, the periodic modulation imposed by the grating on the incident field enforces a far-field diffraction pattern presenting a structure of interference maxima that is also periodic. The condition for interference maxima is satisfied at multiples of the far-field displacement ${x^\prime}=m T'$, $\forall$ $m \in \mathbb{Z}$, with $T'$ given by
\begin{eqnarray} 
\label{Tprime} 
T^\prime= \frac{2\pi}{\xi T}=  f \frac{\lambda}{T}.
\end{eqnarray} 

The periodic pattern of far-field interference may also be understood from the fact that the Fourier transform of a Dirac delta-comb (with period $T$) is, itself, a Dirac delta-comb with the period  $T^\prime$ given by Eq. \eqref{Tprime}: the sum in Eq. \eqref{DiffPer} running with unbounded $n$ and  constant amplitudes $A_n$ $\forall$ $n$ is indeed the Fourier series of a Dirac comb \cite{Lohmann1992}. 
In any experiment, however, only a finite number of slits is in fact illuminated, causing the phasors $e^{-i n T \xi  {x^\prime}}$ from \eqref{DiffPer} to get nearly aligned around the positions of the Dirac comb corresponding to complete constructive interference. 

\section{Results}
\label{sec:Results}

\subsection{Far-field diffraction periodicities}

\begin{figure}[t!]
\begin{center}
\includegraphics[width=84mm]{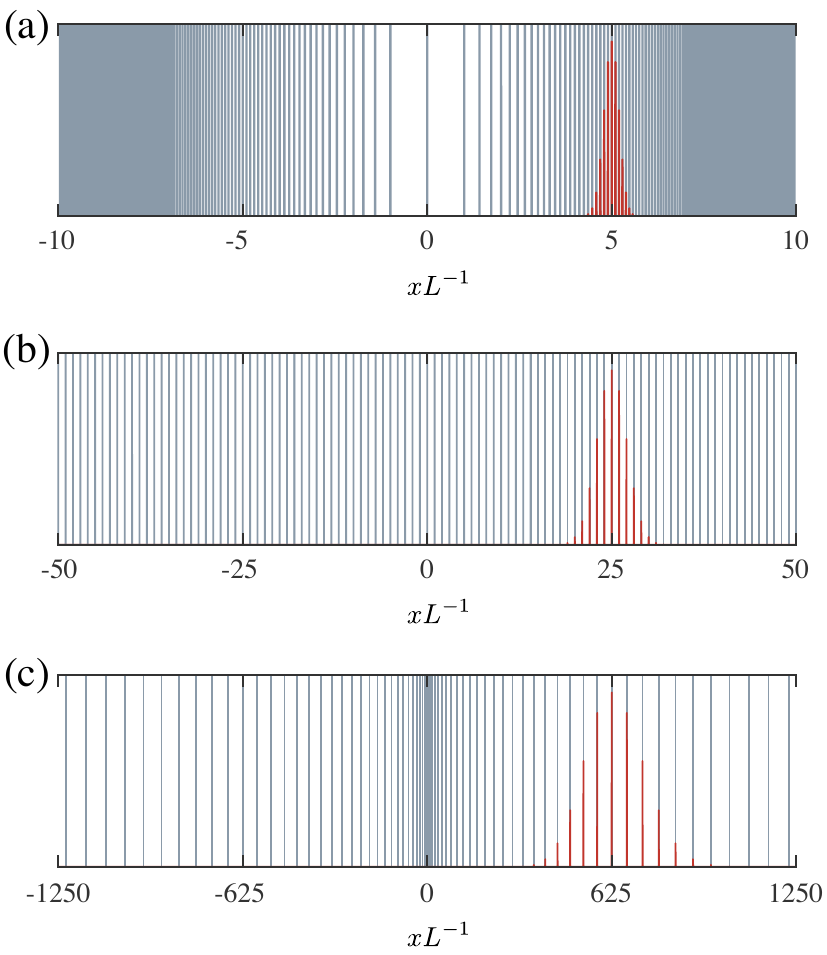}
\caption{Representation of mirror-symmetric slit arrays (grey colour) constructed from the slit position function $q_{\pm n}= \pm n^pL$ [see \eqref{q_n}] using $x_0=0$ [see \eqref{SlitPos}], and (a) $p=1/2$, (b) $p=1$, and (c) $p=2$. For these arrays, we used  $s=8\mu$m as the slit width and $L=500\mu$m. Notice that the abscissa of the graphs refers to the dimensionless coordinate $xL^{-1}$. The red plots represent the intensity of the diffracting field behind the arrays [the absolute square of \eqref{Psigeneral}] using the Gaussian amplitude distribution, \eqref{A_nGauss}, with mean slit number $\bar{n}=25$ and $\sigma_n/\sqrt{2}=2$ representing the standard deviation of $|A_n|^2$. The corresponding far-field diffraction patterns [the absolute square of \eqref{DiffGeneral}] are plotted in Fig. \ref{Fig:FarFieldNumerics}.}
\label{Fig:SlitArray}
\end{center}
\end{figure}

So far, we haven't specified a particular geometry for the slit array. Apart from the case of a periodic grating, treated in Section \ref{subsec:periodic}, the far-field diffraction given in Eq. \eqref{DiffGeneral} can be quite intricate, presenting oscillations of different spatial frequencies that are dependent on the non-linear behaviour of the slit position function $q_n$ appearing in Eq. \eqref{SlitPos}. Of particular interest in this work are non-linear functions exhibiting a power law dependence on the slit number ($p \in \mathbb{R}$)
\begin{equation}  \label{q_n}
q_{\pm n} =\pm  n^p  L, 
\end{equation}
where $L$ is a constant with dimension of length. By construction, these functions exhibit odd parity, enforcing arrays that present mirror-symmetry. In Fig. \ref{Fig:SlitArray} we illustrate three different arrays with vertical grey slits positioned according to Eq. \eqref{q_n} using $p=1/2$ (part a), $p=1$ (part b) and $p=2$ (part c). The intensity of the light field behind the slits [the absolute square of Eq. \eqref{Psigeneral}] is plotted in red. For these plots, the field amplitude distribution $A_n$ was taken as a Gaussian,
\begin{equation}\label{A_nGauss}
A_n\propto \exp\left[-\frac{(n-\bar{n})^2}{2\sigma_n^2}\right],
\end{equation}
with mean $\bar{n}=25$ and $\sigma_n=2\sqrt{2} $. Notice that when plotted against $xL^{-1}$, the intensity distribution is concentrated around the dimensionless value $(\bar{n})^p$ of the abscissa.

\begin{figure}[t!]
\begin{center}
\includegraphics[width=84mm]{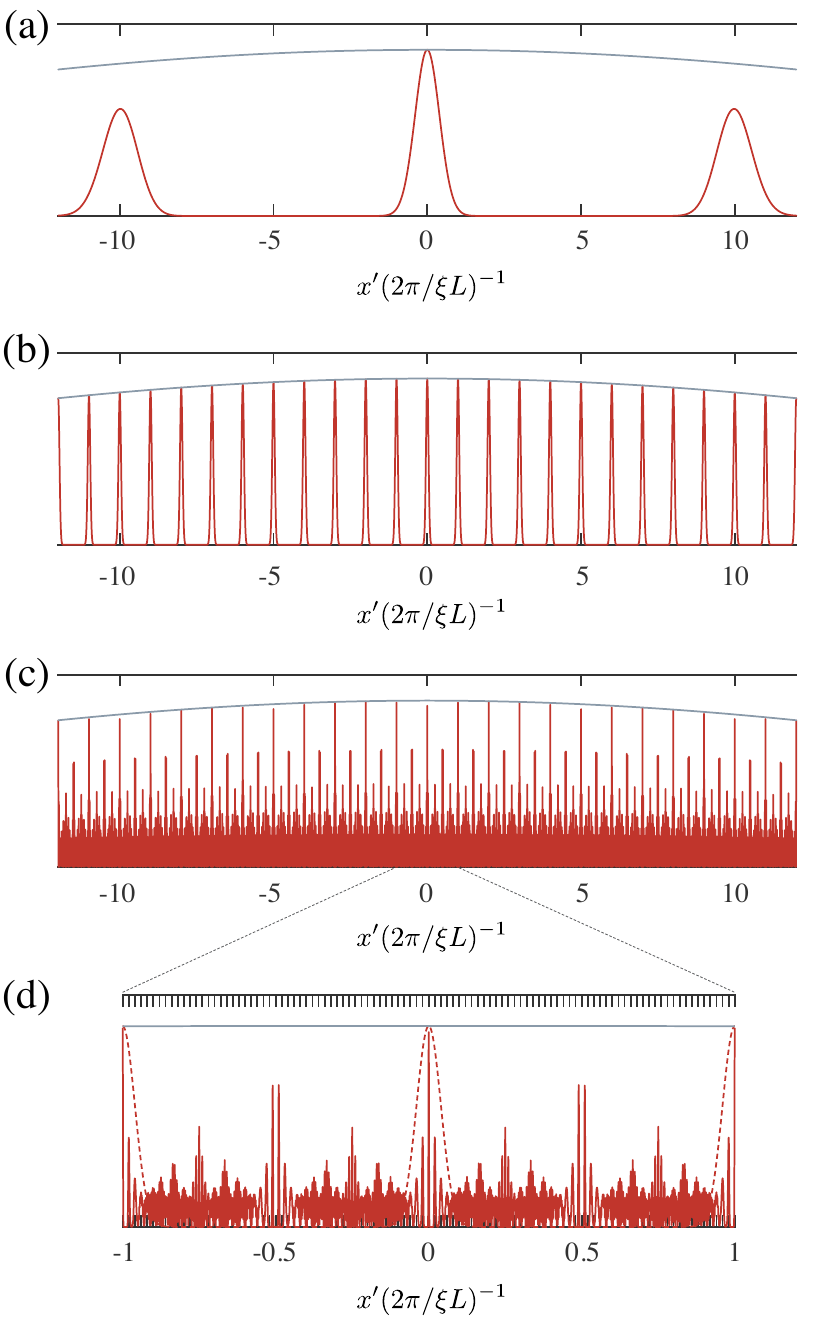}
\caption{Far-field diffraction patterns (red colour) obtained numerically from the diffracting fields represented in Fig. \ref{Fig:SlitArray}. The corresponding slit arrays obey \eqref{q_n} using  (a) $p=1/2$, (b) $p=1$, and (c) $p=2$. The grey curves correspond to the diffraction pattern of a single slit [see \eqref{SlitFourier}]. In units of $T^\prime= {2\pi}/{\xi L}$, the first order periodicities are (a) $10$, (b) 1 and (c) $1/50$. Note, also, that the \textit{second} order periodicity for $p=2$ exactly matches the \textit{first} order periodicity for the periodic array. In part (d) of the figure we plot the far-field pattern associated to $p=2$ (solid red line) in the interval $[-1,1]$ with a $\times 10$ zoom, and ticks spaced 1/50 apart. For comparison, the red dashed line shows the far-field pattern of the periodic array.}
\label{Fig:FarFieldNumerics}
\end{center}
\end{figure}

We shall now use the Taylor expansion  \cite{Robinett04} of Eq. \eqref{q_n} to identify the distance scales at which interference from the aperiodic array may produce diffraction peaks. Looking back upon Eq. \eqref{DiffGeneral}, we recognise the most significant contributing terms for the far-field diffraction as those whose amplitude coefficients $A_n$ are appreciable. A Taylor series expansion of \eqref{q_n} around $\bar{n}$ gives
\begin{equation}  \label{TaylorExpansion}
q_n \approx \sum_{j=0}^{\infty} \frac{q_{\bar{n}}^{(j)}}{j!} (n-\bar{n})^j = q_{\bar{n}}  +  q_{\bar{n}}^{(1)}(n-\bar{n}) +  \frac{q_{\bar{n}}^{(2)}}{2!}(n-\bar{n})^2 + \cdots,
\end{equation} 
where the derivatives are evaluated considering a continuous interpolation of the slit position function: $q_n \rightarrow q(n)$, with $n \in \mathbb{R}$,
\begin{equation}  \label{Derivatives}
q_{\bar{n}}^{(j)} \coloneqq \left. \frac{d^{(j)} q(n) }{dn^{(j)}} \right|_{n=\bar{n}} 
\end{equation} 
Assuming a dispersion $\Delta n$ for the illumination distribution such that $\bar{n} \gg \Delta n \gg 1$, the slit position function is well approximated by the first few terms of the expansion \eqref{TaylorExpansion}, suggesting that we define the following quantities ($j \in \mathbb{N}$) 
\begin{equation}  \label{OscillationPeriods}
T^\prime_{j} \coloneqq  \frac{2\pi }{\xi \left|q_{\bar{n}}^{(j)} \right|}j! = \frac{\lambda f}{\left|q_{\bar{n}}^{(j)}\right|}j!,
\end{equation} 
which we will refer to as the \emph{$j$-th order periodicities} of the far-field diffraction pattern. Indeed, plugging the Taylor expansion \eqref{TaylorExpansion} together with the definitions \eqref{Derivatives} and \eqref{OscillationPeriods} into \eqref{DiffGeneral}, we rewrite the far-field diffraction pattern as
\begin{equation} 
\label{DiffGeneralTaaylor} 
\tilde{\psi}({x^\prime}) = \tilde{\Pi}_0({x^\prime})  e^{-i \frac{2\pi}{T^\prime_{0}}  {x^\prime}} 
 \sum_{n\in \mathbb{Z}} A_n  e^{-i \sum_{j=1}^{\infty} \frac{2\pi}{T^\prime_{j}} (n-\bar{n})^j  {x^\prime} }.
\end{equation} 
Notice that, for $q_{n}$ given by Eq. \eqref{q_n}, these quantities satisfy
\begin{equation}\label{Tratio}
\frac{T^\prime_{j+1}}{T^\prime_{j}} =\frac{j+1}{|p-j|} \bar{n} \gg 1,
\end{equation}
as long as we consider $\bar{n} \gg 1$. When $p$ is an integer, $T^\prime_{j}~\to~\infty$ $\forall$ $j > p$, so strictly speaking expression \eqref{Tratio} holds only for  $j \leq p$.  In other words, each $j$-th order periodicity describes a very different length scale, which increases in value with $j$. As we will see below, these distances correspond precisely to the scales at which revivals of different orders occur in the far-field interference pattern.

\begin{table}[t] 
\begin{center}
\begin{tabular}{c  c  c  c  } 

& {$p=0.5$}
&{$p=1$}
& {$p=2$} \\
\hline \hline

{$T^\prime_0/T^\prime $}   
& $\bar{n}^{-0.5}$
& $\bar{n}^{-1}$
& $\bar{n}^{-2}$  \\ 

{$T^\prime_1 /T^\prime $}	
& $2  \bar{n}^{+0.5}$	
& $1$	
& $\frac{1}{2}\bar{n}^{-1}$	 \\
	
{$T^\prime_2 /T^\prime $}
& $8  \bar{n}^{+1.5}$ 		
& $\rightarrow \infty$	
& $1$    \\

{$T^\prime_3 /T^\prime $}
& $16 \bar{n}^{+2.5}$	
& $ \rightarrow \infty$	
& $ \rightarrow \infty$   \\
\hline \hline

\end{tabular}
\caption{Periodicities of the far-field diffraction pattern given in Eq.~\ref{DiffGeneral}, in units of $T^\prime= {2\pi}/{\xi L}$. The diffracting structure is defined as an array of slits with the slit position function presenting a power law dependence on the slit number: $q_{\pm n}=\pm n^pL$, such as in Eq. \eqref{q_n}. The $0$-th order periodicity, $T^\prime_0$, is only observable in the far-field pattern of mirror-symmetric diffracting fields [see Eqs. \eqref{DiffPlusMinusFinal} and \eqref{FFMS1}].}
\label{TabPeriod}
\end{center}
\end{table}%

It is useful at this point to look at specific examples. For positive integer $p$, the $j$-th order periodicities, \eqref{OscillationPeriods}, are given by ($p \in \mathbb{N}^+$)
\begin{equation}  \label{Derivatives p+}
T^\prime_{j} = \left( \frac{2\pi }{\xi L} \right) j! \frac{|p-j|!}{p!}\bar{n}^{(j-p)},
\end{equation} 
expression which extends to positive real $p$ ($p \in \mathbb{R}^+$) upon substitution of the factorial with the gamma function: $ x! =\Gamma(x+1)$ $\forall$ $x \in \mathbb{N}^+$.
In Table \ref{TabPeriod} we present the periodicities \eqref{OscillationPeriods} associated with the first four terms of the Taylor expansion \eqref{TaylorExpansion} for the particular examples  $p=0.5$, $1$ and $2$. For $p=1$ we recover the periodic grating with the length parameter $L=T$ representing the period of the grating and the first order periodicity $T^\prime_1=\left(2\pi /\xi L\right)=T^\prime$ standing for the periodicity of the far-field diffraction pattern found in Eq. \eqref{Tprime}. Interestingly, the second order periodicity $T^\prime_{2}$ for the case where $p=2$ is also independent of $\bar{n}$, and in fact exactly matches the first order periodicity $T^\prime_{1}$ of the far-field diffraction pattern obtained from a periodic array with $T=L$. Indeed, it is easy to see from Eq. \eqref{Derivatives p+} that this will happen for every $p$, \textit{i.e.} $T^\prime_{j=p} = \left(2\pi /\xi L\right)$  $\forall$ $p \in \mathbb{N}^+$ and $q_{n}$ given by \eqref{q_n}.

We numerically Fourier transform [according to Eq. \eqref{OpticalFourier}] the diffracting fields represented in Fig. \ref{Fig:SlitArray} to obtain the corresponding far-field diffraction patterns, plotted in red in Fig. \ref{Fig:FarFieldNumerics}. The `$\mathrm{sinc}$' envelope corresponding to the diffraction pattern of a single slit [the absolute square of Eq. \eqref{SlitFourier}] is plotted in grey.
To better compare the periodicities obtained from \eqref{OscillationPeriods} with the far-field pattern obtained from numerics, we use the dimensionless coordinate ${x^\prime} (2\pi/\xi L)^{-1}$ as the abscissa of the plots. According to Table \ref{TabPeriod} (using $\bar{n}=25$), the first order periodicity for $p=0.5$ is ten times greater than that of the periodic array, $T'$, whilst the second order periodicity, a thousand times greater than $T'$, predicts a diffraction peak lying well beyond the domain where the amplitude of the `$\mathrm{sinc}$' envelope is appreciable. For $p=2$, the first order periodicity is $1/50$. 

\subsection{Diffraction peaks and modulating envelope for mirror-symmetric fields}
\label{MSfield}

The intensity patterns shown in Fig. \ref{Fig:FarFieldNumerics} do not reveal diffraction peaks at distance scales corresponding to the zeroth order periodicity $T^\prime_0$, summarised in the first row of Table \ref{TabPeriod}. Most commonly, this periodicity is not observable through the far-field \textit{intensity} pattern, as it appears in the argument of a global phase factor in Eq. \eqref{DiffGeneralTaaylor}. 
In order to observe diffraction peaks at distance scales of $T^\prime_0$, we may interfere the diffracting fields represented in Fig. \ref{Fig:SlitArray} with a mirror copy of themselves, in which case the complex amplitude coefficients relate as $A_{n}=A_{-n}$. This even amplitude distribution yields a diffracting field behind the slit array whose reflection symmetry (around the central coordinate $x_0$) is written as $\psi_n(2x_0-x)=\psi_{-n}(x)$.
For simplicity, we shall assume a flat wavefront illumination: $A_{n}=|A_{n}| e^{i\varphi}$, where $\varphi$ does \textit{not} depend on $n$. In this case, the resulting  pattern of far-field diffraction displays Hermitian symmetry, $\tilde{\psi}_n(-{x^\prime})=\tilde{\psi}_n^\ast({x^\prime})$, where `$\ast$' stands for complex conjugation and we neglect the constant phase factor $\varphi$. Under inversion of the slit number, the symmetry of the far-field diffraction pattern is
\begin{equation}
\label{SymmetryFF} 
\tilde{\psi}_{-n}({x^\prime}) =  \tilde{\psi}^\ast_{n}({x^\prime}) e^{-i2\xi x_o {x^\prime}},
\end{equation}
implying that opposite slits generate a far-field diffraction pattern with the same absolute value: $|\tilde{\psi}_{n}({x^\prime})|=|\tilde{\psi}_{-n}({x^\prime})|$. 

We shall now make use of the discussed symmetry properties to express the resulting diffraction pattern. Let us split the complex function, Eq. \eqref{DiffGeneral}, into positive and negative frequency parts, such that
\begin{eqnarray} 
\label{DiffPlusMinus} 
\tilde{\psi}({x^\prime})= \tilde{\Pi}_0({x^\prime})\left[ \Upsilon_+({x^\prime}) + \Upsilon_-({x^\prime}) \right],
\end{eqnarray} 
where
\begin{eqnarray} 
\label{ComplexSum} 
\Upsilon_{\pm}({x^\prime}) \coloneqq \frac{ A_0}{2} +\sum_{\substack{\pm n >0}} A_n e^{-i \kappa_n  {x^\prime}}.
\end{eqnarray} 
It follows that both complex functions have the same absolute value, $\left|\Upsilon_+({x^\prime}) \right| =\left|\Upsilon_-({x^\prime}) \right|$, and can be factored out in Eq. \eqref{DiffPlusMinus}  as 
\begin{equation} \label{DiffPlusMinusFinal} 
\tilde{\psi}({x^\prime})= 2\tilde{\Pi}_0({x^\prime}) \left|\Upsilon_+({x^\prime}) \right|   \cos{[ \theta_+({x^\prime})]},
\end{equation} 
where we define the arguments
\begin{equation} \label{argument} 
\theta_+({x^\prime}) \coloneqq \arg[\Upsilon_{+}({x^\prime})]=-\arg[\Upsilon_{-}({x^\prime})].
\end{equation} 
We thus see that the absolute value of the complex function, the definition \eqref{ComplexSum}, behaves as an \textit{envelope}, modulating the oscillatory interference term. Note, yet, that \eqref{DiffPlusMinusFinal} reduces to the expression of a double slit interference whenever a slit $m$ is illuminated along with its mirror image 
 ($A_n=A_{-n}=A_m\delta_{nm}$), in which case the argument of the oscillatory term in \eqref{argument} is simply $\theta_+({x^\prime})=\kappa_m{x^\prime}$ and the envelope $\left|\Upsilon_+({x^\prime}) \right| =A_m$ is constant. Finally, notice that we can also interpret the diffraction pattern arising from the mirror symmetric field, Eq. \eqref{DiffPlusMinusFinal}, as the interference of a number of double-slit fields:
\begin{equation} \label{FFMS1}
\tilde{\psi}({x^\prime})= \tilde{\Pi}_0({x^\prime}) \left[A_0 + 2 \sum_{n>0}  A_n \cos( \kappa_n  {x^\prime}  )
\right].
\end{equation} 
%

\subsection{Quantum mechanical analogy 1: evolution of localised quantum wave packets} \label{QMA1}

The general treatment of the far-field diffraction produced by slit arrays presented in this paper is in perfect analogy with the well studied phenomenon of \textit{wave packet} revivals \cite{Bluhm96,Robinett04}, according to which an initially \textit{localised} quantum state may return to its initial condition after evolution under a certain Hamiltonian. A quantum state evolving under the Hamiltonian $\hat{H}$ may be written as
\begin{equation} \label{QuantumState}
| \psi(t) \rangle = \sum_{n \in \mathbb{N}} c_n e^{- i \omega_n t} | E_n \rangle,
\end{equation} 
where $\hat{H} |E_n \rangle = E_n|E_n \rangle $, $\omega_n=E_n/ \hbar$ and the coefficients $c_n=\langle E_n | \psi(0) \rangle$ give the amplitude distribution of the initial state in terms of the energy eigenstates. The time evolution of this quantum state is often analysed by means of the \textit{autocorrelation function} \cite{Robinett04}, measuring the overlap of \eqref{QuantumState} with the initial ($t=0$) quantum state 
\begin{equation} 
\label{Autocorrelation} 
\mathcal{C}(t) \coloneqq \langle \psi(0)  | \psi(t) \rangle = \sum_{n} |c_n|^2 e^{-i \omega_n t}.
\end{equation} 
A localised wave packet at $t=0$ may be represented by a Gaussian distribution $|c_n|^2=A_n$ such as that in Eq. \eqref{A_nGauss}, with mean $\bar{n}$ much greater than its standard deviation $\Delta n =\sigma_n$. 
The similarity between Eq. \eqref{DiffGeneral} and the autocorrelation defined above permits us to map $\mathcal{C}(t)$ on to $\tilde{\psi}({x^\prime})$ with the spatial angular frequencies $\kappa_n$ defined in \eqref{kappan} playing the role of $\omega_n$ and the spatial coordinate ${x^\prime}$ playing the role of time. In this analogy, the geometry of the slit array defined through Eq. \eqref{q_n} simulates the Hamiltonian's energy spectrum:
\begin{equation} 
\label{En wavepacket} 
E_n  \rightarrow \hbar \xi q_n = \hbar \frac{2\pi}{ f \lambda} q_n.
\end{equation} 
%

\begin{table}[t] 
\begin{center}
\begin{tabular}{c  c  c } 

Far-field diffraction & Localised QWP& Atomic inversion JCM\\
\hline \hline
$\tilde{\psi}({x^\prime})$ - Eq. \eqref{DiffGeneral}& $\mathcal{C}(t)$ -   Eq. \eqref{Autocorrelation}  & $\langle \hat{\sigma}_z \rangle (t)$ - Eq. \eqref{AtomicInversion}    \\
\hline

$\kappa_n$	& $\omega_n $		& $\Omega_{(n-1)}$ \\   
$A_n$		& $|c_n|^2	$		& $2|c_{(n-1)}|^2$ \\   
$T^\prime_0$		& $t_{\mathrm{ZB}}$	& $t_{\mathrm{R}}$ \\   
$T^\prime_1$		& $t_{\mathrm{cl}}$	&  $t_{\mathrm{R,r} }$ \\
$T^\prime_2$		& $t_{\mathrm{r}}$	& $t_{\mathrm{R,sr} }$ \\
$T^\prime_3$		& $t_{\mathrm{sr}}$	& -- \\

\hline \hline
\end{tabular}
\caption{Analogy between the pattern of far-field diffraction and the functions representing the autocorrelation of localised quantum wave packets (QWP) and the atomic inversion in the Jaynes-Cummings Model (JCM). The subindices of the evolution time scales stand for \textit{zitterbewegung} (ZB), classical (cl), revival (r), super-revival (sr), Rabi (R), Rabi revival (R,r) and Rabi super-revival (R,sr).}
\label{TabAnalogy}
\end{center}
\end{table}%

The slit position function of a \textit{periodic} array ($q_n \propto n$) is identified with the spectrum of the quantum harmonic oscillator ($E_n \propto n$), whereas the \textit{aperiodic} array with $p=2$ reproduces a spectrum proportional to the square of the quantum number ($E_n \propto n^2$), such as that of the infinite potential well \cite{Bluhm96,Robinett04}. The intensity diffraction patterns shown in part (d) of Fig. \ref{Fig:FarFieldNumerics} numerically reproduce $|\mathcal{C}(t)|^2$ for quantum systems evolving under these two Hamiltonians. The first order periodicity of the far-field diffraction pattern, $T^\prime_1$, represents the \textit{classical} evolution periodicity of the quantum state. For example, for the harmonic oscillator, the quantum evolution is exactly periodic for any initial condition - corresponding to an exactly periodic far-field interference pattern in the optical setup. The period $T^\prime_1 = T'$ is just the classical oscillator period. For more general aperiodic spectra, it can be seen from Eqs.~\eqref{Derivatives} and \eqref{OscillationPeriods} that $T^\prime_1/h$ will be of the order of the inverse of the smallest energy gap between relevant energy levels (i.e., those with $n \sim \bar{n}$). In this case the evolution is only quasiperiodic, and after a few multiples of $T^\prime_1$ have passed the various frequency components of the quantum state will have lost their initial phase relationships.

The second order periodicity, $T^\prime_2$,  represents evolution times for which the most important of these components approximately recover their initial phase relationships. At these moments the state therefore returns, approximately, to its initial condition: $\mathcal{C}(mt_{\mathrm{r}}) \approx 1$, where $m \in \mathbb{N}$ and $t_{\mathrm{r}}$ is the \textit{revival} time \cite{Bluhm96,Robinett04}. We thus see that revivals in the context of the time evolution of localised wave packets are a consequence of the quantised nature of the Hamiltonian's spectrum, $E_n$. Note that, as shown in Eq.~(\ref{Tratio}), such revivals generally occur at times much larger than the classical period. An extreme example of this occurs for the case of the linear spectrum (harmonic oscillator). In this case, as noted above, the evolution is already perfectly periodic at the first order period, so there are no revivals to speak of ($t_r \rightarrow \infty$). This is illustrated for example by the fact that coherent Gaussian wavepackets (corresponding to coherent states satisfying $\hat{a}|\alpha\rangle = \alpha|\alpha\rangle$) maintain their shape while oscillating back and forth.

Higher order periodicities such as $T^\prime_3$ are analogous to \textit{super-revival} times, $t_{sr}$, for which the whole structure of \textit{lower order} periodicities is reproduced \cite{Dutra:1994,Fang:2000}. On the other hand, according to \eqref{Tratio}, the lowest order periodicity $T^\prime_0$ produces the shortest pattern of diffraction peaks. As seen in subsection \ref{MSfield}, this rapid oscillatory pattern results from the diffraction of mirror-symmetric fields. Localised quantum wave packets evolving under non-linear Hamiltonians with energy spectrum presenting this kind of symmetry ($E_{\pm n} \propto \pm \sqrt{n}$) have been shown to reproduce oscillations at this time scale, $t_{ZB}$, with `ZB' standing for  \textit{zitterbewegung} \cite{Romera2009,Krueckl2009,Garcia2014}. In table \ref{TabAnalogy} we summarise the discussed analogy.

\subsection{Quantum mechanical analogy 2: expectation values of observables \& the Jaynes-Cummings model}\label{QMA2}

The Jaynes-Cummings model (JCM) \cite{Greentree2013} is a simple model for light-matter interaction, coupling a single quantised radiation mode $|n\rangle$ to a two-level `atom' with ground and excited states $|g\rangle$  and $|e\rangle$, respectively. In its simplest form (assuming atom and electromagnetic radiation are resonant, and making the so-called rotating-wave approximation \cite{Zeuch2020}), it can be described, in the interaction picture, by the Hamiltonian $\hat{H}_{\mathrm{JCM}}=\gamma(\hat{a}^\dagger \hat{\sigma}_- + \hat{a} \hat{\sigma}_+)$, where $\hat{a}$, $\hat{a}^\dagger$ are the annihilation/creation operators for the radiation mode and $\hat{\sigma}_{\pm}\coloneqq (\hat{\sigma}_x\pm i \hat{\sigma}_y)/2$ are the atomic raising/lowering operators written in terms of the Pauli Matrices ($\hat{\sigma}_i$, $i= x, y, z$). The constant $\gamma$ represents the atom-field coupling strength.

A well-known property of this model is the occurrence of so-called \textit{collapses} and \textit{revivals} in quantities such as the expectation value of the atomic inversion: the observable $\hat{\sigma}_z \coloneqq |e\rangle \langle e| - |g\rangle \langle g|$ that measures the degree to which the atom is polarised towards the excited or ground states. For an initial state of the form $|\psi(0)\rangle=|e\rangle \otimes | \phi_{\mathrm {field}} \rangle$, where $| \phi_{\mathrm {field}} \rangle =\sum_{n\in \mathbb{N}}c_n |n\rangle$ represents an arbitrary state of the quantised field with photon number probabilities $P_n=|c_n|^2$, the expectation value of the atomic inversion evolves as
\begin{equation} 
\label{AtomicInversion} 
\langle \hat{\sigma}_z \rangle (t)= \langle \psi(0)  | \psi(t) \rangle = \sum_{n} |c_n|^2 \cos({\Omega_n t}),
\end{equation} 
where
\begin{equation} 
\label{RabiFreq} 
\Omega_n = 2 \gamma \sqrt{n+1},
\end{equation} 
are known as the \textit{Rabi frequencies}. The field state $| \phi_{\mathrm {field}} \rangle$  interacting with the atom (inside a cavity) is often considered to be coherent (an attenuated laser beam). In this case, the probabilities $p_{n}$ for different photon numbers $n$ are described by a Poisson distribution with mean $\bar{n}$, given by
\begin{equation} 
\label{Poisson} 
P_n= \left| c_n \right|^2
= \left| \langle n| \phi_{\mathrm {field}} \rangle \right|^2 
= e^{-\bar{n}}\frac{(\bar{n})^n}{n!}.
\end{equation} 

It is possible to give an expression that \textit{approximates} the structure of the atomic inversion $\langle \hat{\sigma}_z \rangle (t)$ around the midpoint of each revival \cite{{Fleischhauer1993,Jonathan99}}. For the \textit{first} revival, this is given by: 
\begin{equation} 
\label{PeakRevival} 
\langle \hat{\sigma}_z \rangle (t) \approx  \frac{ \gamma t}{\sqrt{2}\pi}
P \left[ n= \left(\frac{\gamma t}{2\pi}\right)^2 \right] 
\cos \left[  \frac{(\gamma t)^2}{2\pi}  -\frac{\pi}{4}\right],
\end{equation} 
where  $P(n)$ is a continuous interpolation of the Poisson distribution in Eq. \eqref{Poisson}. 

The expectation value of the observable $\hat{\sigma}_z$ evolving in time, Eq. \eqref{AtomicInversion}, bears close resemblance to the spatial distribution of far-field diffraction given in Eq. \eqref{FFMS1}. Using $p=1/2$ in the slit position function, Eq. \eqref{q_n}, we can map the Rabi frequencies $\Omega_n$ on to the  spatial angular frequencies given in Eq. \eqref{kappan} as
\begin{equation} 
\label{Omegan JCM} 
\Omega_n  \rightarrow \kappa_{(n+1)}= \xi q_{(n+1)} = \xi L \sqrt{n+1},
\end{equation} 
with the coupling strength constant $\gamma$ mapped as
\begin{equation} 
\label{gamma map} 
\gamma  \rightarrow \frac{\xi L}{2}= \frac{ \pi L}{\lambda f}.
\end{equation} 
In table \ref{TabAnalogy} we summarise the discussed analogy for the JCM and define the relevant periodicities associated with the time evolution of the atomic inversion given in Eq. \eqref{AtomicInversion}: the periodicity of the Rabi oscillations, $t_{R}$, its revival, $t_{R,r}$, and super-revival. In the next section, we shall describe the experimental implementation of Eq. \eqref{AtomicInversion} through the far-field diffraction pattern.

\section{Experimental implementation}
\label{sec:Experiment}

\begin{figure*}[t!]
\begin{center}
\includegraphics[width=168mm]{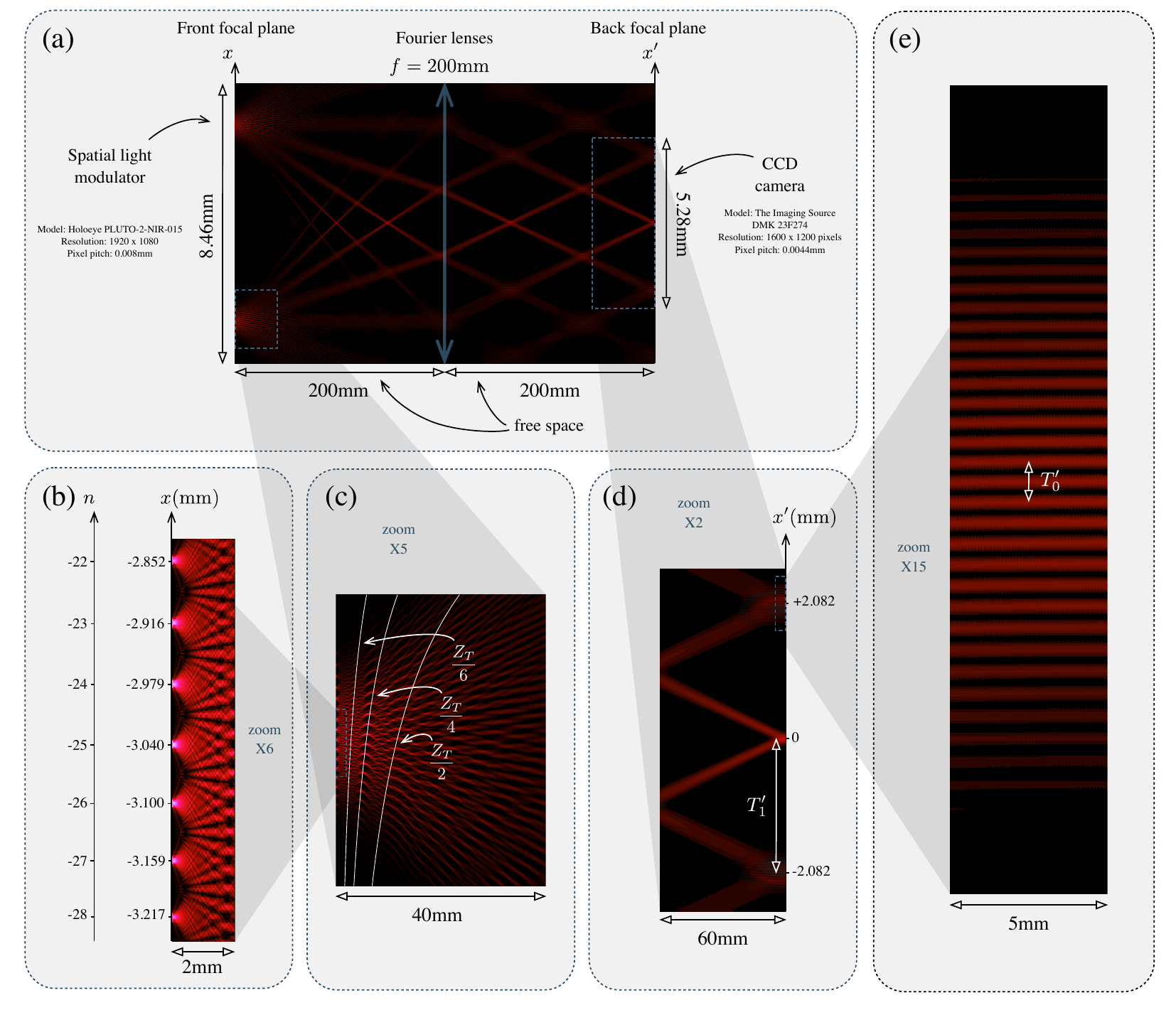} 
\caption{Illustration of the optical setup used to compute patterns of quantum revivals via Fraunhofer diffraction. 
(a) Layout of our experiment using a SLM and a CCD camera at the front and back focal planes of an optical Fourier transform system with 
$\xi\approx49.6$mm$^{-2}$ [see Eq. \eqref{OpticalFourier}]. 
Part (b) zooms into the front focal plane (around $\bar{n}=25$) where the He-Ne laser diffracts  from an aperiodic array of slits following \eqref{q_n} with $p=0.5$ and $L=(0.608\pm 0.008)$mm. 
The near field propagation from the aperiodic array generates a distorted Talbot-like carpet with geometry reproduced by \eqref{TalbotDist} shown in part (c).
The images shown in parts (d) and (e) zoom into the back focal plane to show the generated Fraunhofer diffraction pattern. For this illustration with $\bar{n}=25$, the predicted far-field periodicities are $T^\prime_0=0.042$mm and $T^\prime_1=2.082$mm.
\textit{Numerical resolution:} the field intensity is numerically calculated with resolution of $0.001$mm for the \textit{transverse} and $0.05$mm for the \textit{longitudinal} coordinates; for plot (b) we improved the resolution to account for propagation steps of $0.001$mm.
All images in this figure were plotted using the colormap scheme reported by A. D. Green in Ref. \cite{Green11}.}
\label{Fig:Exp}
\end{center}
\end{figure*}

The optical computation of the quantum revivals discussed in section \ref{sec:Results} is illustrated in Fig. \ref{Fig:Exp}. At the front focal plane of a Fourier transforming lens, we place a spatial light modulator (SLM) able to generate a programmable diffracting field, Eq. \eqref{Psigeneral}, from a collimated He-Ne laser ($\lambda=632.8$nm) impinged on it. The prepared field is Fourier transformed (effective focal length $f=200$mm) according to \eqref{OpticalFourier} and its Fraunhofer diffraction pattern registered on a CCD camera at the back focal plane. The illustration of our experimental procedure shown in part (a) of Fig. \ref{Fig:Exp} was performed numerically based on the Fresnel quadratic phase dispersion formula for the field angular spectrum \cite{goodman_1996} and a thin lens (see the figure caption for details). 

We chose an aperiodic array of slits based on Eq. \eqref{q_n} with $p=0.5$, $L=(0.608 \pm 0.008)$mm and slit width $s=0.008$mm corresponding to a \textit{single} pixel of our SLM. The field amplitudes $A_n$ were programmed using well known amplitude modulation techniques \cite{Davis1999,Clark16} as the Poisson distribution given in Eq. \eqref{Poisson} with five different mean slit numbers ($\bar{n}=20,25,30,35,40$). For the experiment illustration shown in Fig. \ref{Fig:Exp} [see part (b)] we used $\bar{n}=25$ as the mean number. Our choice of the length parameter $L$ allows us to generate slits up to $n \approx \pm 50$ across of the full height of our SLM, just enough to illuminate the slits of interest (around the chosen $\bar{n}$), thus minimising errors of slit positioning introduced by pixelation.
We thus generate a mirror-symmetric optical field at the SLM written as 
\begin{equation}
\label{PsiExperiment} 
\psi(x)=\psi_+(x)+\psi_-(x),
\end{equation} 
where 
\begin{equation} 
\label{PsiExperiment+} 
\psi_\pm(x) \propto \sum_{n >0} 
\left( e^{-\bar{n}}\frac{(\bar{n})^n}{n!} \right)
\Pi(x \mp L\sqrt{n}),
\end{equation} 
with the function $\Pi(x)$ defined in \eqref{Slit}. 

Parts (b) and (c) of Fig. \ref{Fig:Exp} zoom into the near-field of the aperiodic array (around $x_{-25}$) to see the propagation of $\psi_-(x)$ according to Fresnel's diffraction formula. Interestingly, the \textit{aperiodicity} of the slit array generates a near-field diffraction pattern resembling a \textit{distorted} Talbot carpet. The Talbot distance defined as $Z_T=T^2/\lambda$ gives the propagation distance for which the diffracted field reproduces the intensity pattern from a periodic grating \cite{talbot1836lxxvi,Wen2013}. In our case, the distance between consecutive slits is \textit{not} constant but can be written as a function of the slit index. Defining the \textit{moving average}  $T(x) \coloneqq\left( x_{n+1}-x_{n-1} \right)/2$ as a continuous function of $n$, we plot the white curves on the top of the carpet, corresponding to a Talbot distance that is a function of the transverse coordinate 
\begin{equation} 
\label{TalbotDist} 
Z_T(x)= \frac{ T^2 (x) }{\lambda}.
\end{equation} 
The curves plotted according to Eq. \eqref{TalbotDist} reproduce very nicely the geometry of the distorted Talbot carpet around the slits $n=\pm\bar{n}$.

Parts (d) and (e) of Fig. \ref{Fig:Exp} zoom into the far-field of the aperiodic array. Most interestingly, the observed Fraunhofer diffraction pattern displays periodic structures of interference maxima despite the fact that the diffracting array is not periodic. The first order periodicity $T^\prime_1$ [indicated in part (d)] corresponds to the diffraction of the fields $\psi_+$ or $\psi_-$ alone, whereas the zeroth order periodicity $T^\prime_0$ [indicated in part (e)] is the result of the far-field interference between them.

\begin{figure*}[t!]
\begin{center}
\includegraphics[width=168mm]{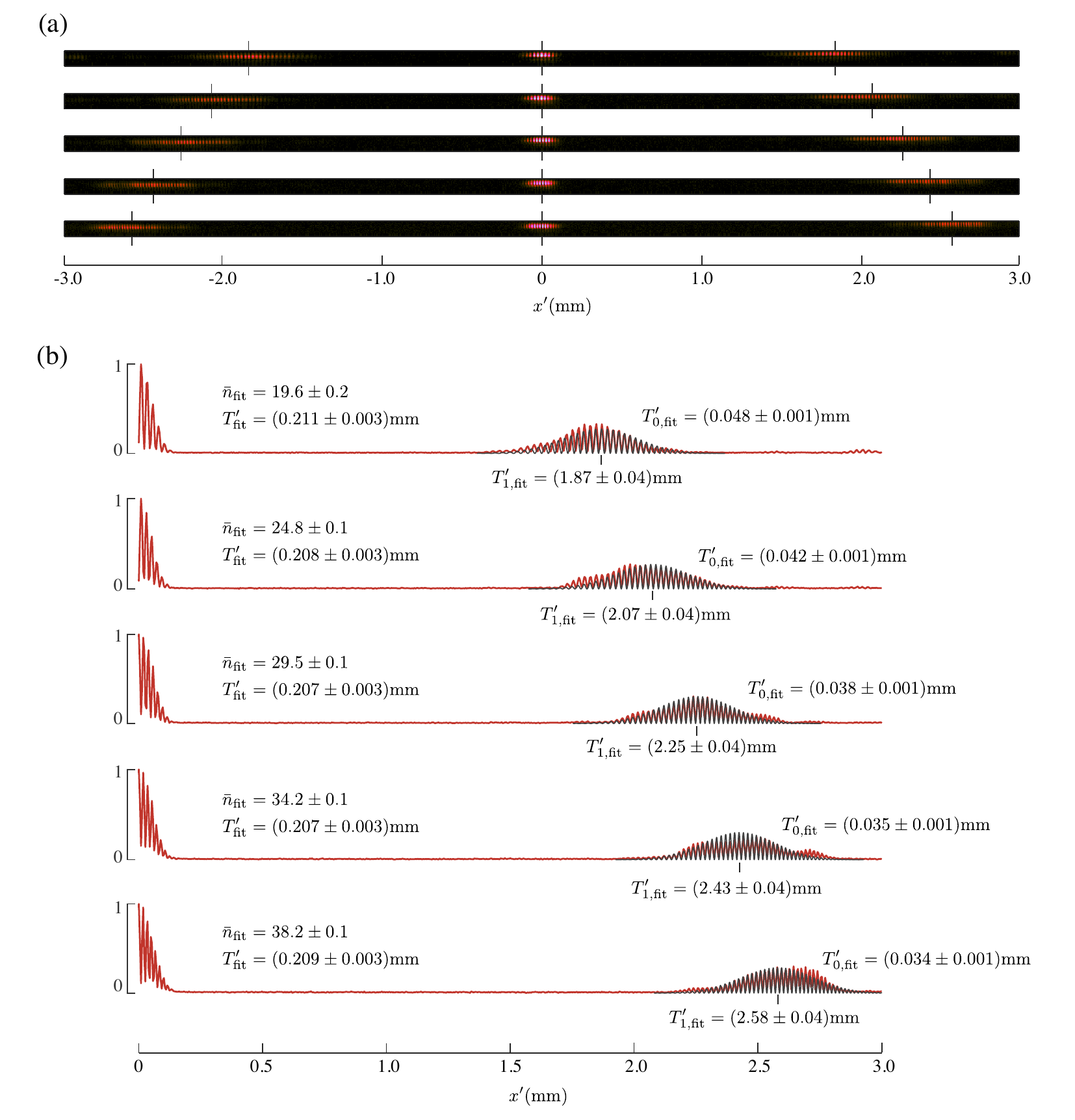} 
\caption{Experimental results obtained for the optical computations of the atomic inversion given in \eqref{AtomicInversion}. (a) Images of the far-field diffraction patterns registered on the CCD camera. (b) The red plots were obtained by summing the corresponding images across the 21 pixels in the vertical direction, subtracting the background and normalising them to their maximum values. The first revival of each measurement is then fitted using the absolute square of \eqref{PeakRevival}. The obtained fitted values of $T^\prime_{\mathrm{fit}}$ and $\bar{n}_{\mathrm{fit}}$ are used to plot the black curves. Notice that the periodicity of the diffraction peaks of interference maxima is \textit{one half} of the zeroth order periodicity, since we are here considering the \textit{intensity} of the diffracted light field.}
\label{Fig:ExpResults}
\end{center}
\end{figure*}

In Fig. \ref{Fig:ExpResults} we present the experimental results obtained for the optical computation of the quantum revivals illustrated numerically in Fig. \ref{Fig:Exp}. The far-field diffraction patterns registered on the CCD camera display periodic modulations of interference maxima at two distinct distance scales. The rapid oscillatory pattern of interference maxima and their revivals are dependent on the mean slit number chosen as parameter for the field amplitude at the front focal plane, Eqs. \eqref{PsiExperiment} and \eqref{PsiExperiment+}. In order to compare these two periodicities in a consistent  manner with those summarised in table \ref{TabPeriod}, we make use of the approximate expression for the structure of the first revival in the atomic inversion, Eq. \eqref{PeakRevival}. We first sum the recorded images (21 pixels across the CCD camera), subtract the background (the minimum intensity value) and normalise the resulting intensities to their maximum values, as plotted (red colour) in part (b) of Fig. \ref{Fig:ExpResults}. These plots are then used as optical computations of the absolute square of Eq. \eqref{AtomicInversion}. Making use of table \ref{TabAnalogy}, we adapt Eq. \eqref{PeakRevival}  to the far-field amplitude around the coordinate ${x^\prime}=T^\prime_1$ by identifying $T^\prime \leftrightarrow \pi/\gamma$.
 Finally, we fit our normalised intensity data with the absolute square of Eq. \eqref{PeakRevival}, using $T^\prime_{\mathrm{fit}}$ as the \textit{fitted length parameter} and $\bar{n}_{\mathrm{fit}}$ as the \textit{fitted mean} of the Poisson distribution. The obtained fitted quantities (listed in Fig. \ref{Fig:ExpResults}) are used as inputs to Eq. \eqref{PeakRevival} to generate the corresponding revival functions, plotted in black. From the fitted quantities we then calculate the zeroth and first order periodicities of our measured diffraction patterns:  $T^\prime_{0,\mathrm{fit}}=T^\prime_{\mathrm{fit}}({\bar{n}_{\mathrm{fit}}})^{-0.5}$ and $T^\prime_{1,\mathrm{fit}}=T^\prime_{\mathrm{fit}}2({\bar{n}_{\mathrm{fit}}})^{0.5}$.

The approximate expression for the first revival, Eq. \eqref{PeakRevival}, fits very nicely our experimental data: around the coordinate corresponding to the first revival of the rapid oscillatory pattern of interference maxima, the Fourier transform of Eq. \eqref{PsiExperiment} is well described by Eq. \eqref{PeakRevival}.
The values obtained for the fitted length parameter stand very close to that predicted from the experimental parameters, $T^\prime =  \left(2\pi /\xi L\right) = (0.208\pm 0.003)$mm, where the error comes from the uncertainty in $L$ given by the SLM pixel size. This gives a coupling strength constant  $\gamma \rightarrow \pi/T^\prime = (15.1\pm0.2)$mm$^{-1}$ according to our quantum mechanical analogy [see Eq. \eqref{gamma map}] .

Most interestingly, the whole \textit{measured} diffraction patterns lie entirely within the first diffraction order imposed by the slit width. We are then able to modulate the diffraction peaks into a rapid oscillatory pattern that collapses and revives in a very distinct distance scale. For instance, the ratio between the first and zeroth order periodicities [see Eq. \eqref{Tratio}] obtained in our experiment lie between $40 < 2\bar{n}_{\mathrm{fit}} < 80$. Taking the second measurement as an example ($\bar{n}$ programmed as $25$, giving $\bar{n}_{\mathrm{fit}}=24.8$), we obtain approximately 48 peaks of diffraction maxima per milimetre, whereas the spatial frequency of their revival ($\approx 0.48/$mm) is two orders of magnitude smaller. Also interesting to notice is that the revival structure spreads the diffraction peaks of interference maxima across a greater region in space (when compared to the central portion of the diffraction pattern), whilst the intensity of the peaks is decreased. This modulation is not due to the diffraction orders imposed by the slit width and is well understood in the context of quantum revivals \cite{pimenta16}.

\section{Conclusion}
\label{sec:conclusion}
We have shown how to structure light into interesting patterns of quantum revivals using aperiodic arrays and Fraunhofer diffraction. The key features of recurrence phenomena in the evolution of quantum systems are the quantisation of the energy levels and the \textit{non-linear} dependence of the energy spectrum on the quantum number. In our work we implement these features with aperiodic gratings defined by non-linear functions obeying a power law dependence on the position of the slits in the array. 

Revivals, or recurrence phenomena, have been observed on a number of systems. For instance, the oscillations of a carbon nanotube subject to forces that deviate from the linear relation on the displacement, \textit{i.e.}, a deviation from Hookes's Law, have been shown to evolve and revive at certain time intervals \cite{Barnard19}. Recurrence phenomena due to the paraxial propagation of light waves known as self-imaging have been recently measured \cite{Silva2020}. In another interesting investigation, the long-term evolution of electronic wave packets described by Rydberg states was experimentally observed \cite{Yeazell1990}. 
This latter case can be easily implemented in our optical setup using aperiodic arrays with $p=-2$ in Eq. \eqref{q_n}. Our optical method is then an important contribution for the better understanding of quantum revivals and can be used to compute different patterns of time evolution in recurrence phenomena. 

Moreover, we have numerically shown that aperiodic arrays generate a near-field structure of interference maxima well described by distorted Talbot-like carpets. For periodic gratings, this near-field diffraction generates a periodic structure of interference maxima, the Talbot carpet, a recurrence phenomenon experimentally demonstrated both using optical \cite{Case:2009} as well as the matter \cite{Chapman:1995} waves. The relation between periodicities of conjugate variables distributions has important consequences on unbiasedness relations \cite{Tasca18a} and entanglement detection \cite{Tasca:2018b} of continuous variables quantum systems submitted to coarse-graining. Our investigation on quasiperiodic arrays and their interference structures may have important consequences in these topics.

\textbf{Funding.}
Air Force Office of Scientific Research (FA9550-19-1-0361); Instituto Nacional de Ci\^encia e Tecnologia de Informa\c{c}\~ao Qu\^antica (465469/2014-0); Conselho Nacional de Desenvolvimento Cient\'ifico e Tecnol\'ogico (431804/2018-4, 304332/2018-6); Funda\c{c}\~ao Carlos Chagas Filho de Amparo \`a Pesquisa do Estado do Rio de Janeiro (E-26/201.414/2021, E-26/201.108/2021).

\textbf{Acknowledgments.}
We would like to thank E. F. Galv\~ao, G. B. Lemos, R. de Melo e Souza and S. P. Walborn for interesting discussions and feedback on our work. D.S.T. acknowledges Leonardo Calv\~ao for useful feedback on graphics design for the preparation of the figures. 

\textbf{Disclosures.}
The authors declare no conflicts of interest.

\textbf{Data availability.} Data underlying the results presented in this paper are not publicly available but may be obtained from the authors upon reasonable request.

%


\end{document}